\documentclass[11pt]{article}
\usepackage{enumerate}
\usepackage{amsmath}
\usepackage{amssymb}
\usepackage{bm}
\usepackage{ascmac}
\usepackage{amsthm}
\usepackage{braket}
\usepackage{tikz}
\usetikzlibrary{calc}

\usepackage{a4wide}
\usepackage{cases}
\usepackage{mathrsfs}
\usepackage{authblk}

{\theoremstyle{plain}%
  \newtheorem{thm}{\bf Theorem}[section]%
  \newtheorem{lem}[thm]{\bf Lemma}%
  \newtheorem{cor}[thm]{\bf Corollary}%
}
{\theoremstyle{remark}
  
}

\newcommand{\dC}{{\mathbb{C}}}%
\newcommand{\dN}{{\mathbb{N}}}%
\newcommand{\dZ}{{\mathbb{Z}}}%
\newcommand{\dQ}{{\mathbb{Q}}}%

\title{Odd-periodic Grover walks}
\author[2]{Yusuke YOSHIE\footnote{Corresponding author, E-mail: y-yoshie@ishikawa-nct.ac.jp}}
\affil[2]{General Education, Ishikawa College, National Institute of Technology, Ishikawa 929-0392, Japan}\date{}
\begin {document}
\maketitle{}
\begin{abstract}
The Grover walk is one of the most well-studied quantum walks on graphs. In this paper, we investigate its periodicity to reveal the relationship between the quantum walk and the underlying graph, focusing particularly on the characterization of graphs exhibiting a periodic Grover walk. Graphs having a periodic Grover walk with periods of $2, 3, 4$, and $5$ have previously been characterized. It is expected that graphs exhibiting a periodic Grover walk with odd period correspond to cycles with odd length. We address that problem and are able to perfectly characterize the class of graphs exhibiting an odd-periodic Grover walk by using a combinatorial method.   
 \end{abstract}

\section{Introduction}
A quantum walk is a quantum analogue of a random walk \cite{Aharonov, Gudder}, and it has been applied to several research fields, including graph theory, quantum physics, and computer science. Details can be found in \cite{Emms, Wang, Portugal, venegas, Watrous}. Quantum walks on graphs are currently in the limelight because of their ability to show us specific characteristics that are not seen in classical random walks, and these specific characteristics are the focus of this paper. Specifically, we address the {\it periodicity} of a discrete-time quantum walk. If there exists a positive integer $k$ such that the time evolution $U$ satisfies $U^{k}=I$, where $I$ is the identity operator, then the quantum walk is periodic. In other words, any initial state $\varphi_{0}$ returns to itself after $k$ applications of the time evolution. If there exists such an integer, then the minimum one is called the {\it period}. Nowadays, research on the periodicity of discrete-time quantum walks is a topic of much interest, especially its ability to characterize graphs exhibiting a periodic quantum walk. Konno {\it et al.}\cite{Konno17} considered the periodicity of the Hadamard walk in cycles and was able to determine conditions that have periodic Hadamard walks. Additionally, Higuchi {\it et al.} \cite{Higuchi13} discussed periodicities of discrete-time quantum walks in several graphs with high symmetry, including complete graphs, complete bipartite graphs, strongly regular graphs, and cycles. Moreover, Saito \cite{Saito} formulated the Fourier walk on cycles and addressed the periodicity of it. Kubota {\it et al.}\cite{SETY2} proposed the staggered walk on a generalized line graph induced by a Hoffman graph and addressed its periodicity. 

In this paper, we study the periodicity of the {\it Grover walk}, which is strongly related to the structure of the underlying graph. We call a graph exhibiting a periodic Grover walk a {\it periodic graph}. The Grover walk is only defined by the underlying graph, and it strongly reflects the graph structure. If a graph is periodic, then its structure is sometimes restricted. For example, elements of the graph structure, such as the degree and diameter, are restricted if the Grover walk on the graph is periodic. Such restriction is seen in \cite{SETY, Yoshie2}. Thus, the Grover walk is a suitable model for revealing the relation between graph structure and its induced quantum walk. 
Yoshie \cite{Yoshie} studied its periodicity and found classes of graphs exhibiting a periodic Grover walk with a given period. The results for periodic graphs are summarized in the following table, where $P_{n}$ and $C_{n}$ are the path graph on $n$ vertices and the cycle graph on $n$ vertices, respectively.
\begin{table}[h]
\begin{center}
\begin{tabular}{|c|c|} \hline
Period & Graph \\ \hline
2 & $P_{2}$ \\ \hline
3 & $C_{3}$ \\ \hline
4 & every complete bipartite graph \\ \hline
5 & $C_{5}$ \\ \hline
over 6 & still open \\ \hline
\end{tabular}
\end{center}
\end{table}
The researchers in \cite{Yoshie} studied the characterization of periodic graphs with odd period. Using a relatively easy calculation, they demonstrated that every cycle graph with odd length induces a periodic Grover walk of odd period. Hence, being a cycle graph with odd length is a sufficient condition for inducing an odd-periodic Grover walk. 
However, demonstrating its necessity is not so easy. It is thought that cycle graphs with odd length are the only ones that admit an odd-periodic Grover walk (necessity holds). We address this problem in this paper and show that the statement is true. Specifically, we show that an odd-periodic graph with period $k$ is the only cycle graph with length $k$ for an odd $k$. This is formally stated in the following theorem. 
\begin{thm}
Let $k$ be an odd integer. A graph $G$ is periodic with period $k$ if and only if $G=C_{k}$. 
\label{odd period} 
\end{thm} 
Therefore, we completely characterize periodic graphs with an odd period. This statement implies that the period of almost every periodic graph is even.  

%

The remainder of this paper is organized as follows. In Section 2, we give definitions of graphs and the Grover walk and introduce the periodicity. Additionally, we give the transition matrix of the graphs, which plays an important role in this paper. Using them, we prepare spectral tools: eigenvalues and the coefficient of the characteristic polynomial of the transition matrix, for example. After that, we state the main theorem of the paper. In Section 3, we prove our main theorem. Using the tools we prepared, we control the inner structure of graphs and bring them closer to a cycle with odd length. Section 4 is devoted to the summary of our work and to discussion of our future research goals.

\section{Preliminaries}
\subsection{Graph}
In this paper, we only treat simple, connected, and finite graphs. Let $G=(V, E)$ be a graph with vertex set $V$ and edge set $E$. If two vertices $u$ and $v$ are adjacent by an edge, then we denote this as $u \sim v$. For $uv \in E$, we denote the arc from $u$ to $v$ as $(u,v)$. Moreover, the origin and the terminus of $e=(u,v)$ are denoted by $o(e)$ and $t(e)$, respectively. We denote by $\bar{e}$ the inverse arc of $e$. Define $\mathcal{A}:=\{(u,v), (v,u) \mid uv \in E\}$. 

A graph containing no odd cycle as a subgraph is called a bipartite graph. Furthermore, an {\it odd unicycle graph} is a non-bipartite graph satisfying $|V|=|E|$. In other words, an odd unicycle graph is a graph containing exactly one odd cycle as a subgraph. The length of the cycle is called the {\it girth}. For $k \in \dN$, we define $\mathscr{O}(k)$ to be the class of odd unicycle graphs of girth $k$. Note that the cycle graph $C_{k}$ is a member of $\mathscr{O}(k)$. Additionally, a graph containing no cycle as a subgraph is called a {\it tree}. Furthermore, the set of disjoint edges of $G$ is called a {\it matching}. If the number of edges of a matching is $k$, then it is called a {\it $k$-matching}. Throughout this paper, we denote the $n \times n$ identity matrix as $I_{n}$.  

\subsection{Grover walk and its periodicity}
In this section, we define the time evolution operator of the Grover walk and consider its periodicity.
For a graph $G$, let us define $U=\{U_{e,f}\}$ on $\dC^{|\mathcal{A}|}$ by
\begin{equation}
U_{e, f}=
	\begin{cases}
	2/\deg{t(f)}, & \text{if $t(f)=o(e), e\neq \bar{f}$}\\
        2/\deg{t(f)}-1, & \text{if $e=\bar{f}$}\\
        0, & \text{otherwise.}
	\end{cases}
	\label{def of u}
\end{equation}  

Let $\varphi_{0} \in \dC^{|\mathcal{A}|}$ be the initial state. Then the state at time $k$, $\varphi_{k}$, is given by $\varphi_{k}=U^{k}\varphi_{0}$. If there exists $k \in \dN$ such that 
\begin{equation}
U^{k}=I_{|\mathcal{A}|}, 
\label{period}
\end{equation}
then the Grover walk is {\it periodic}. If the Grover walk on a graph $G$ is periodic, then the minimum integer satisfying the condition in (\ref{period}) is called the {\it period}. If the period is specified as $k$, then we call the graph a {\it $k$-periodic graph}, but if the period of the Grover walk induced by a periodic graph $G$ is odd (resp. even), then we call $G$ an odd-(resp. even-) periodic graph. 



\subsection{Transition matrix and its characteristic polynomial}
Let us define the transition matrix $T=\{T_{u,v}\}$ on $\dC^{V}$ induced by a graph $G$ as
\begin{equation}
T_{u, v}=
\begin{cases}
\frac{1}{\deg{u}}, & \text{if $u \sim v$}\\
0, & \text{otherwise},
\end{cases}
\end{equation}
which is a weighted adjacency matrix expressing the isotropic random walk on $G$. Note that 
\[ Tf(v)=\sum_{u \sim v}\frac{1}{\deg{v}}f(u) \]
for $f \in \dC^{|V|}$, where $f(u)$ is the entry of $f$ corresponding to $u$. Let $\mathrm{Spec}(\cdot)$ be the set of eigenvalues. We can now present a spectral mapping theorem for the Grover walk. 
\begin{thm}[Higuchi and Segawa \cite{HSegawa17}]
Let $U$ be defined as above. Then the set of eigenvalues of $U$ can be expressed as 
\[\mathrm{Spec}(U)=\{ e^{\pm i \cos^{-1}{ \left( \mathrm{Spec}(T) \right)} } \}  \cup {\{ 1 \}}  \cup  {\{ -1 \} }.  \]
\label{spec decom}
\end{thm}
Let $\lambda_{1}, \dots, \lambda_{n}$ be the eigenvalues of $U$. If it holds that there exists $k_{i} \in \dN$ such that
\[ \lambda^{k_{i}}_{i}=1 \]
and
\begin{eqnarray*}
\lambda^{j}_{i} \ne 1, &  j < k_{i} 
\end{eqnarray*}
for $1 \le i \le n$, then $G$ is periodic and the period is given by
\begin{equation}
\mathrm{lcm}(k_{1}, \dots, k_{n}), 
\label{lcm}
\end{equation}
where $\mathrm{lcm}(\cdot, \dots, \cdot)$ is the least common multiple. If $G$ is $k$-periodic, then it follows from Theorem \ref{spec decom} that $\lambda^{k}=1$ for any $\lambda \in \mathrm{Spec}(U)$. From $\{ e^{\pm i \cos^{-1}{ \left( \mathrm{Spec}(T) \right)} }\}$, we obtain the following. 
\begin{cor}
Let $T$ be defined as above. A graph $G$ is periodic if and only if
\[ \cos^{-1}{ \left( \mathrm{Spec}(T) \right)}  \subset \pi \dQ. \]
\end{cor}

We now introduce another tool for investigating the periodicity. Let $n=|V|$, and let $\rho_{j}$ be the coefficient of the characteristic polynomial of $T$ for $0 \le j \le n$, that is,
\[\det{(xI_{n}-T)}=\sum^{n}_{j=0}\rho_{j}x^{j}.\]
\begin{lem}[Yoshie \cite{Yoshie2}]
If $G$ is periodic, then 
\begin{equation}
2^{j}\rho_{n-j} \in \dZ 
\end{equation}
\label{lem of rho}
for $0 \le j \le n$. 
\end{lem}
Recall the definition of the determinant. Let $Y=xI_{n}-T$ and $V(G)=\{v_{1}, v_{2}, \dots, v_{n}\}$. Then it holds that
\begin{equation}
\det{Y}= \sum_{\sigma}\mathrm{sgn}(\sigma)Y_{v_{1}, \sigma(v_{1})} Y_{v_{2}, \sigma(v_{2})}\dots Y_{v_{n},\sigma(v_{n})},
\label{def of det}
\end{equation}
where $\sigma$ runs over the permutations on $V(G)$, and $\mathrm{sgn}(\cdot)$ is the signature. We call permutations like
\[ \sigma=\left(
\begin{array}{cccc}
v_{1} & v_{2} & \dots & v_{r}\\
v_{2} & v_{3} & \dots & v_{1}
\end{array}
\right)
\] 
and
\[ \sigma^{-1}=\left(
\begin{array}{cccc}
v_{1} & v_{2} & \dots & v_{r}\\
v_{r} & v_{1} & \dots & v_{r-1}
\end{array}
\right)
\] 
{\it cyclic permutations} with length $r$. $\sigma$ maps $v_{i}$ to $v_{i+1}$ for $1 \le i \le r-1$ and $v_{r}$ to $v_{1}$. Additionally, we denote by $|\sigma|$ the length of $\sigma$, and a permutation with length $2$ is called a {\it transposition}. Note that 
\[
Y_{v_{i}, \sigma(v_{i})}=
\begin{cases}
x, & \text{if $v_{i}=\sigma(v_{i})$}\\
-\frac{1}{\deg{v_{i}}}, & \text{if $v_{i} \sim \sigma(v_{i})$}\\
0, & \text{otherwise.}
\end{cases}
\]
Thus, the permutations contributing to the coefficient of $x^{n-j}$ in (\ref{def of det}) are expressed by a product of cyclic permutations $\sigma_{1}, \sigma_{2}, \dots, \sigma_{k}$ so that $|\sigma_{1}|+|\sigma_{2}|+\dots+|\sigma_{k}|=j$. Hence, a permutation like
\[ \sigma=\left(
\begin{array}{cccc}
v^{(1)}_{1} & v^{(1)}_{2} & \dots & v^{(1)}_{j_{1}}\\ 
v^{(1)}_{2} & v^{(1)}_{3} & \dots & v^{(1)}_{1}\\ 
\end{array}
\right)\cdot
\left(
\begin{array}{cccc}
v^{(2)}_{1} & v^{(2)}_{2} & \dots & v^{(2)}_{j_{2}}\\ 
v^{(2)}_{2} & v^{(2)}_{3} & \dots & v^{(2)}_{1}\\ 
\end{array}
\right)\cdots
\left(
\begin{array}{cccc}
v^{(k)}_{1} & v^{(k)}_{2} & \dots & v^{(k)}_{j_{k}}\\ 
v^{(k)}_{2} & v^{(k)}_{3} & \dots & v^{(k)}_{1}\\ 
\end{array}
\right)
\] 
with $\{v^{(i)}_{1}, v^{(i)}_{2}, \dots, v^{(i)}_{j_{i}}\} \cap \{v^{(l)}_{1}, v^{(l)}_{2}, \dots, v^{(l)}_{j_{l}}\}=\phi$ for $i \ne l$ and $v^{(i)}_{1} \sim v^{(i)}_{2} \sim \dots \sim v^{(i)}_{j_{i}} \sim v^{(i)}_{1}$, and $j_{1}+j_{2}+\dots j_{k}=j$ contributes the term
\begin{equation} 
\mathrm{sgn}(\sigma)\prod^{k}_{i=1}\prod^{j_{i}}_{s=1}\left( -\frac{1}{\deg{v^{(i)}_{s}}}\right)
\label{sum}
\end{equation}
to the coefficient of $x^{n-j}$. Thus, $\rho_{n-j}$ is obtained by summing (\ref{sum}) over all permutation with length $j$. A cyclic permutation and a transposition correspond to a cycle and an edge in the underlying graph. Thus, computing the coefficient of the characteristic polynomial is equivalent to finding a combination of these cycles and matchings. 


\section{Proof of Theorem \ref{odd period}}
\subsection{Non-odd-periodic graph}
Here, we present a non-odd-periodic graph that plays an important role. To this end, let us introduce a {\it Chebyshev polynomial}. We inductively define a sequence of polynomials $\{U_{n}(x)\}^{\infty}_{n=0}$ through the following process: $U_{0}(x)=1$, $U_{1}(x)=2x$, and
\begin{equation}
U_{n+1}(x)=2x U_{n}(x)-U_{n-1}(x) 
\label{rec}
\end{equation}
for $n \ge 1$. This is the Chebyshev polynomial of the second kind. The following is a well-known property of the Chebyshev polynomial:
\[ U_{n}(\cos{\theta})=\frac{\sin(n+1)\theta}{\sin{\theta}}. \]
It is convenient to set $U_{-1}(x)=-1$. More detailed properties of the Chebyshev polynomials can be found in \cite{Riv}.

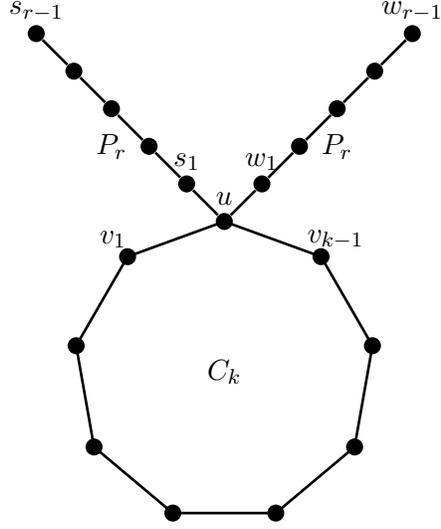
\begin{figure}[h]
\begin{center}
\begin{tikzpicture}
[scale = 1.0,
fat/.style={circle,fill=black, inner sep = 2mm},
slim/.style={circle,fill=black, inner sep = 0.8mm},
]
\foreach \angle /\label in
{0,1,..., 8}
{
\node[slim] (v\angle) at (90+40*\angle:2) {};
\draw[line width=1pt] (90+40*\angle:2) -- (90+40*\angle+40:2);
} 
\node at (0,0) {$C_{k}$};
\node at (0,2.3) {$u$};
\node at (1.5, 3) {$P_{r}$};
\node at (-1.5, 3) {$P_{r}$};
\node at (-0.5,2.8) {$s_{1}$};
\node at (-2.5, 4.8) {$s_{r-1}$};
\node at (0.5,2.8) {$w_{1}$};
\node at (2.5, 4.8) {$w_{r-1}$};
\node at (90+40:2.3) {$v_{1}$};
\node at (90-40:2.3) {$v_{k-1}$};

\node[slim] (x0) at (0.5,2.5) {};
\node[slim] (x1) at (1,3) {};
\node[slim] (x2) at (1.5,3.5) {};
\node[slim] (x3) at (2,4) {};
\node[slim] (x4) at (2.5,4.5) {};
\node[slim] (y0) at (-0.5,2.5) {};
\node[slim] (y1) at (-1,3) {};
\node[slim] (y2) at (-1.5,3.5) {};
\node[slim] (y3) at (-2,4) {};
\node[slim] (y4) at (-2.5,4.5) {};
\draw[line width=1pt] (v0)--(x0)--(x1)--(x2)--(x3)--(x4);
\draw[line width=1pt] (v0)--(y0)--(y1)--(y2)--(y3)--(y4);
\end{tikzpicture}
\caption{A non-odd-periodic graph}
\label{even graph}
\end{center}
\end{figure}

We construct a graph $G$ by connecting an odd cycle $C_{k}$ and two path graphs $P_{r}$ with the same length. Now, we identify a vertex in the cycle $C_{k}$ and the endpoints of two paths $P_{r}$ as a single vertex $u$. Let $V(G)=\{u, v_{1}, \dots, v_{k-1}, s_{1}, \dots, s_{r-1}, w_{1}, \dots, w_{r-1}\}$ and 
\[ E(G)=\{ v_{i-1}v_{i} \mid 1 \le i \le k\} \cup \{ s_{i-1}s_{i} \mid 1 \le i \le r-1\} \cup \{ w_{i-1}w_{i} \mid 1 \le i \le r-1\}, \]
where $v_{0}=v_{k}=s_{0}=w_{0}=u$ (see Figure \ref{even graph}). 
\begin{thm}
The graph illustrated in Figure \ref{even graph} is not an odd-periodic graph.
\label{even period}
\end{thm}
\begin{proof}
For $1 \le l \le r-1$, set $\lambda_{l}=\cos{\frac{2l-1}{2(r-1)}\pi}$. Define $f_{l} \in \dC^{|V|}$ as 
\begin{align*}
f_{l}(s_{j})&=U_{j-1}(\lambda_{l}),\\
f_{l}(w_{j})&=-U_{j-1}(\lambda_{l}),\\
f_{l}(v_{i})&=0
\end{align*}
for $1 \le i \le k$ and $1 \le j \le r-1$. We show that $f_{l}$ is an eigenvector of $T$ associated with $\lambda_{l}$. First, it clearly holds that $Tf_{l}(v_{i})=\lambda_{l}f_{l}(v_{i})$ for $1 \le i \le k-1$. Next, we have
\begin{align*}
Tf_{l}(u)&=\frac{1}{4}(f_{l}(v_{1})+f_{l}(v_{k-1})+f_{l}(s_{1})+f_{l}(w_{1}))\\
&=\frac{1}{4}(0+0+U_{0}(\lambda_{l})-U_{0}(\lambda_{l}))\\
&=0=\lambda_{l}f_{l}(u). 
\end{align*}
Moreover, it holds that 
\begin{align*}
Tf_{l}(s_{j})&=\frac{1}{2}(f_{l}(s_{j+1})+f_{l}(s_{j-1}))\\
&=\frac{1}{2}(U_{j}(\lambda_{l})+U_{j-2}(\lambda_{l}))\\
&=\frac{1}{2}(2\lambda_{l}U_{j-1}(\lambda_{l})-U_{j-2}(\lambda_{l})+U_{j-2}(\lambda_{l}))\\
&=\lambda_{l}U_{j-1}(\lambda_{l})=\lambda_{l}f_{l}(s_{j})
\end{align*}
for $1 \le j \le r-2$. Setting $j=r-1$, we have
\begin{align*}
Tf_{l}(s_{r-1})&=\frac{1}{1}f_{l}(s_{r-2})\\
&=U_{r-3}(\lambda_{l})\\
&=\frac{\sin{(r-2)\frac{2l-1}{2(r-1)}\pi}}{\sin{\frac{2l-1}{2(r-1)}\pi}}\\
&=\lambda_{l}U_{r-2}(\lambda_{l})=\lambda_{l}f_{l}(s_{r-1}). 
\end{align*}
Similarly, it holds that $Tf_{l}(w_{j})=\lambda_{l}f_{l}(w_{j})$ for $1 \le j \le r-1$. Thus, it holds that $Tf_{l}=\lambda_{l}f_{l}$, so $f_{l}$ is an eigenvector associated with $\lambda_{l}$. Then
\[ \left\{ \cos{\frac{2l-1}{2(r-1)}\pi} \mid 1 \le l \le r-1 \right\} \subset \mathrm{Spec}(T). \]
Note that $\cos{\frac{2l-1}{2(r-1)}\pi}$ is the real part of a $4(r-1)$-th root of unity. By (\ref{lcm}), the period is a multiple of $4(r-1)$ if $G$ is periodic. Therefore, $G$ is not odd-periodic. 
\end{proof}

\subsection{Necessary condition for graphs to be odd-periodic}
A necessary condition for graphs to be odd-periodic can be found in \cite{Yoshie}. 
\begin{thm}[Yoshie \cite{Yoshie}]
If $G$ is odd-periodic, then $G$ is an odd unicycle graph. 
\label{odd cycle}
\end{thm} We want to obtain a stronger condition for necessity. Suppose that a graph $G$ is odd-periodic. It follows from Theorem \ref{odd cycle} that $G$ is an odd unicycle graph. In other words, $G \in \mathscr{O}(k)$ for an odd integer $k \ge 3$. Define $u_{0}, u_{1}, \dots, u_{k-1}$ as the vertices of the unique cycle in $G$. This is the unique combination of vertices satisfying $u_{0} \sim u_{1} \sim \dots \sim u_{k-1} \sim u_{0}$. First, we prove the following statement.
\begin{thm}
If $G \in \mathscr{O}(k)$ for an odd $k$ is periodic, then one of the following holds:
\begin{itemize}
\item[(i)] $\deg{u_{0}}=\deg{u_{1}}=\dots=\deg{u_{k-1}}=2$
\item[(ii)] there exits $0 \le i \le k-1$ such that $\deg{u_{i}}=4$ and $\deg{u_{l}}=2$ for $l \ne i$.
\end{itemize}
\label{thm girth}
\end{thm}
\begin{proof}
We begin by analyzing the coefficient of the characteristic polynomial of the transition matrix $T$. Let 
\[ \det{(xI_{n}-T)}=\sum^{n}_{j=0}\rho_{j}x^{j}, \]
where $n=|V|$. Then the only permutations that contribute $\rho_{n-j}$ are ones like 
\[ \sigma=
\left(
\begin{array}{cccc}
u_{0} & u_{1} & \dots & u_{k-1}\\
u_{1} & u_{2} & \dots & u_{0}
\end{array}
\right)\]
and $\sigma^{-1}$ since $k$ is odd and $\{u_{0}, u_{1}, \dots, u_{k-1}\}$ is the unique set of vertices satisfying $u_{0} \sim u_{1} \sim \dots \sim u_{k-1} \sim u_{0}$. Note that $\mathrm{sgn}(\sigma)=\mathrm{sgn}(\sigma^{-1})=1$ since $k$ is odd. By (\ref{sum}), we have
\[ \rho_{n-k}=\mathrm{sgn}(\sigma)\prod^{k-1}_{i=0}\left( -\frac{1}{\deg{u_{i}}}\right)+\mathrm{sgn}(\sigma^{-1})\prod^{k-1}_{i=0}\left( -\frac{1}{\deg{u_{i}}}\right)=-2\prod^{k-1}_{i=0}\frac{1}{\deg{u_{i}}}. \] 
For $G$ to be periodic, it must hold that $2^{k}\rho_{n-k} \in \dZ$ by Lemma \ref{lem of rho}. Then
\[ -2^{k+1}\prod^{k-1}_{i=0}\frac{1}{\deg{u_{i}}} \in \dZ. \]
Since $\deg{u_{i}} \ge 2$, we have that either $\deg{u_{i}}=2$ for $0 \le i \le k-1$ or there uniquely exists $0 \le j \le k-1$ such that $\deg{u_{j}}=4$ and $\deg{u_{l}}=2$ for $l \ne j$. Thus we complete the proof. 
\end{proof}
If an odd unicycle graph $G$ satisfies (i) in Theorem \ref{thm girth}, then $G$ is nothing but $C_{k}$, which is known to be a $k$-periodic graph \cite{Yoshie}. From now on, we only treat odd unicycle graphs satisfying condition (ii). Our aim is to show that periodic odd unicycle graphs satisfying (ii) are the only ones illustrated in Figure \ref{even graph}. However, the graph is not odd-periodic.  

\subsection{Key statements}
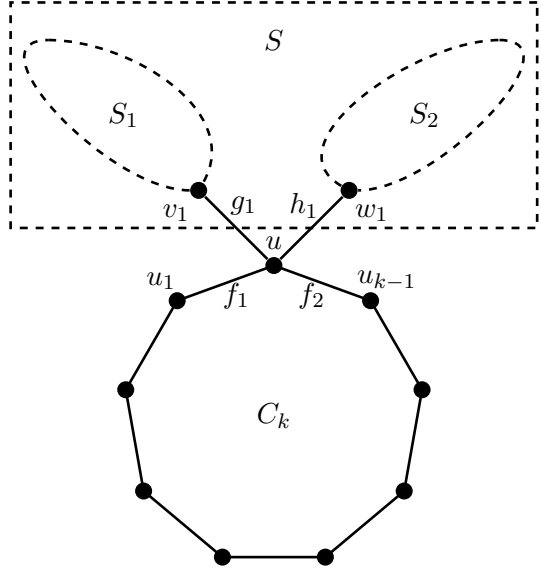
\begin{figure}[h]
\begin{center}
\begin{tikzpicture}
[scale = 1.0,
fat/.style={circle,fill=black, inner sep = 2mm},
slim/.style={circle,fill=black, inner sep = 0.8mm},
]
\foreach \angle /\label in
{0,1,..., 8}
{
\node[slim] (v\angle) at (90+40*\angle:2) {};
\draw[line width=1pt] (90+40*\angle:2) -- (90+40*\angle+40:2);
} 
\node at (0,0) {$C_{k}$};
\node[slim] (x1) at (1,3) {};
\node[slim] (x2) at (-1,3) {};
\draw[line width=1pt] (v0)--(x1);
\draw[line width=1pt] (v0)--(x2);
\draw[line width=1pt, dashed] (x1) to [in=180, out=150] (3,5);
\draw[line width=1pt, dashed] (3,5) to [in=0, out=0] (x1);
\node at (2,4) {$S_{2}$};
\node at (1.3,2.7) {$w_{1}$};

\draw[line width=1pt, dashed] (x2) to [in=180, out=180] (-3,5);
\draw[line width=1pt, dashed] (-3,5) to [in=60, out=0] (x2);
\node at (-2,4) {$S_{1}$};
\node at (-1.3,2.7) {$v_{1}$};
\node at (0,2.3) {$u$};
\node at (1.5,1.8) {$u_{k-1}$};
\node at (-1.5,1.8) {$u_{1}$};

\node at (-0.4,2.8) {$g_{1}$};
\node at (0.4,2.8) {$h_{1}$};
\node at (-0.5,1.6) {$f_{1}$};
\node at (0.5,1.6) {$f_{2}$};

\draw[line width=1pt, dashed] (3.5, 5.5)--(-3.5, 5.5)--(-3.5,2.5)--(3.5,2.5)--(3.5,5.5);
\node at (0,5) {$S$};

\end{tikzpicture}
\caption{The shape of an odd-periodic graph.}
\label{odd graph}
\end{center}
\end{figure}
Suppose that $G \in \mathscr{O}(k)$ is odd-periodic and satisfies condition (ii) in Theorem \ref{thm girth}. 
Let the vertices of the unique cycle $C_{k}$ be $\{u, u_{1}, \dots, u_{k-1} \}$, where $u$ is the unique vertex of degree $4$. Hence, there exist two vertices $v_{1}, w_{1} \not\in V(C_{k})$ such that 
$v_{1} \sim u$ and $w_{1} \sim u$. Since $G$ is a unicycle graph, the subgraph induced by $V(G)\setminus V(C_{k})$, say $S$, does not contain cycles. We decompose $S$ into two subgraphs $S_{1}$ and $S_{2}$ and denote some edges by $f_{1}=uu_{1}, f_{2}=uu_{k-1}, g_{1}=uv_{1}$, and $h_{1}=uw_{1}$ (see Figure \ref{odd graph}). Let $E_{u}=\{ f_{1}, f_{2}, g_{1}, h_{1}\}$ and define $G'$ as the subgraph of $G$ given by
\begin{align*}
V(G')&=V(C_{k})\cup \{ v_{1}, w_{1} \}\\
E(G')&=E(C_{k}) \cup \{ g_{1}, h_{1}\}.
\end{align*}
Furthermore, for $e=xy \in E(G)$, we define a map $M: E(G) \to \dQ$ by
\[ M(e)=\frac{1}{\deg{x}} \frac{1}{\deg{y}}. \]
We also consider the coefficient of the characteristic polynomial of $T$. Throughout this paper, we denote by $\sum_{e_{1}, \dots, e_{l} \in E(S)}$ the summation over the combinations of $l$ edges in $S$ that form an $l$-matching in $E(S)$. Let 
\[ K_{2l}=\sum_{e_{1}, \dots, e_{l} \in E(S)} \prod^{l}_{i=1}M(e_{i}). \]
If there is no such matching in $E(S)$, then we define $K_{2l}=0$. 

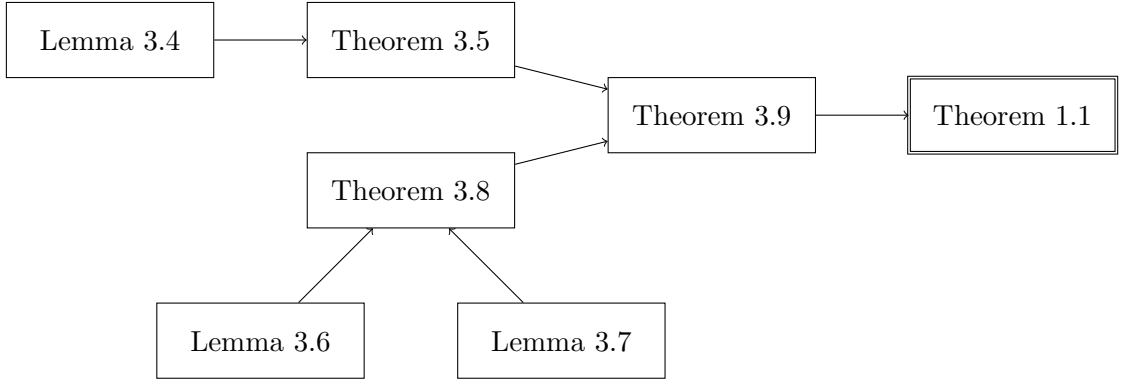
\begin{figure}[h]
\begin{center}
\begin{tikzpicture}
    \tikzset{Terminal/.style={rounded rectangle,  draw,  text centered, text width=3cm, minimum height=1.5cm}};
    \tikzset{Process/.style={rectangle, draw,  text centered, text width=2.5cm, minimum height=1cm}};
     \tikzset{DProcess/.style={rectangle, double, draw,  text centered, text width=2.5cm, minimum height=1cm}};   
    \node[Process](a)at (8,0){Theorem 3.9};
    \node[Process](b)at (4,1){Theorem 3.5};
    \node[Process](c)at (4,-1){Theorem 3.8};
    \node[Process](d) at (0,1){Lemma 3.4};
    \node[Process](e) at (2,-3){Lemma 3.6};
    \node[Process](f)at (6,-3){Lemma 3.7};
    \node[DProcess](g) at (12,0){Theorem 1.1};
     \draw[->] (b)--(a);
     \draw[->] (c)--(a);  
     \draw[->] (d)--(b);
     \draw[->] (e)--(c);
     \draw[->] (f)--(c);
     \draw[->] (a)--(g);
\end{tikzpicture}
\caption{Outline of the proof.}
\label{outline}
\end{center}
\end{figure}

We will now give an outline of the proof. The crucial statement to prove is that the periodic odd unicycle graph in Figure \ref{even graph} can only be the one given by Theorem \ref{main thm2}. To demonstrate this main theorem, we first prove Theorems \ref{deg2} and \ref{main thm1} using Lemmas \ref{lem of K}, \ref{K and L}, and \ref{lem of Kt} (see Figure \ref{outline}). 

\begin{lem}
Let $\rho_{n-k-2t}$ be defined as above for $t \in \dN$ with $k+2t \le n$. If $2^{k+2t}\rho_{n-k-2t} \in \dZ$, then
\[ 2^{2t}K_{2t} \in \dZ. \]  
\label{lem of K}
\end{lem}
\begin{proof}
The permutations contributing $\rho_{n-k-2t}$ are the ones with length $k+2t$ and decomposed into cyclic permutations or transpositions. Since $k+2t$ is odd, the permutations are expressed by the product of a cyclic permutation on $V(C_{k})$ and $t$ disjoint transpositions in $V(S)$, which correspond to a $t$-matching in $E(S)$. The cyclic permutations are
\[\sigma=
\left(
\begin{array}{cccc}
u & u_{1} & \dots & u_{k-1}\\
u_{1} & u_{2} & \dots & u
\end{array}
\right)
\]  
and 
\[\sigma^{-1}=
\left(
\begin{array}{cccc}
u & u_{1} & \dots & u_{k-1}\\
u_{k-1} & u & \dots & u_{k-2}
\end{array}
\right).
\]  
Then we have 
\begin{equation}
\mathrm{sgn}(\sigma)\prod^{k-1}_{i=0}\left(-\frac{1}{\deg{u_{i}}}\right)=\mathrm{sgn}(\sigma^{-1})\prod^{k-1}_{i=0}\left(-\frac{1}{\deg{u_{i}}}\right)=-\frac{1}{2^{k+1}},
\label{pm1}
\end{equation}
where $u_{0}=u$.  Let $\{ x_{1}y_{1}, x_{2}y_{2}, \dots, x_{t}y_{t} \} \subset E(S)$ be a $t$-matching in $E(S)$ and $e_{i}=x_{i}y_{i}$. The transposition on $\{x_{i}, y_{i}\}$ is
\[ \tau=\left(\begin{array}{cc}
x_{i} & y_{i} \\
y_{i} & x_{i}
\end{array}
\right)
\]
and
\[ \mathrm{sgn}(\tau)\left(-\frac{1}{\deg{x_{i}}}\right)\left(-\frac{1}{\deg{y_{i}}}\right)=-M(e_{i}). \]
Then all the $t$-matchings in $E(S)$ yield
\begin{equation}
\sum_{e_{1}, e_{2}, \dots, e_{t} \in E(S)} (-1)^{t} \prod^{t}_{i=1}M(e_{i}). 
\label{pm2}
\end{equation}
Inserting  (\ref{pm1}) and (\ref{pm2}) into (\ref{sum}), we have
\begin{align*}
\rho_{n-k-2t}&=2\left( -\prod^{k-1}_{i=0}\frac{1}{\deg{u_{i}}} \right) \cdot \left(\sum_{e_{1}, e_{2}, \dots, e_{t} \in E(S)} (-1)^{t} \prod^{t}_{i=1}M(e_{i}) \right)\\
&=(-1)^{t+1}\frac{1}{2^{k}}\sum_{e_{1}, e_{2}, \dots, e_{t} \in E(S)} \prod^{t}_{i=1}M(e_{i}).
\end{align*}
The condition $2^{k+2t}\rho_{n-k-2t} \in \dZ$ implies 
\begin{align*}
2^{k+2t} \left( (-1)^{t+1}\frac{1}{2^{k}}\sum_{e_{1}, e_{2}, \dots, e_{t} \in E(S)} \prod^{t}_{i=1}M(e_{i})\right)&=(-1)^{t+1}2^{2t}\sum_{e_{1}, e_{2}, \dots, e_{t} \in E(S)} \prod^{t}_{i=1}M(e_{i})\\
&=(-1)^{t+1}2^{2t}K_{2t} \in \dZ,
\end{align*}
which completes the proof. 
\end{proof}

\begin{thm}
Let $v_{1}$ and $w_{1}$ be defined as above. If $G$ is periodic, then it holds that $\deg{v_{1}}=\deg{w_{1}}=1$ or $2$. 
\label{deg2}
\end{thm}
\begin{proof}
Suppose that a graph $G$ illustrated in Figure \ref{odd graph} is periodic. Then it follows from Lemma \ref{lem of rho} that $2^{k+2}\rho_{n-k-2} \in \dZ$. By Lemma \ref{lem of K}, this condition is reduced to $2^{2}K_{2} \in \dZ$. It holds that 
\[ K_{2}=\sum_{e \in E(S) }M(e)=\sum_{e \in E(G)}M(e)-\sum_{e \in E(G')}M(e). \]
Note that $\sum_{e \in E(G)}M(e)=-\rho_{n-2}$ by (\ref{sum}). Then we have
\begin{align*}
K_{2}&=\sum_{e \in E(G)}M(e)-\sum_{e \in E(G')}M(e)\\
&=-\rho_{n-2}-\left\{\sum^{k-2}_{i=1}M(u_{i}u_{i+1})+M(f_{1})+M(f_{2})+M(g_{1})+M(h_{1})\right\}\\
&=-\rho_{n-2}-\frac{1}{4}(k-2)-\frac{1}{4}-\frac{1}{4\deg{v_{1}}}-\frac{1}{4\deg{w_{1}}}.
\end{align*}
The condition $2^{2}K_{2} \in \dZ$ gives rise to
\[ -2^{2}\rho_{n-2}-(k-2)-1-\left( \frac{1}{\deg{v_{1}}}+\frac{1}{\deg{w_{1}}}\right) \in \dZ. \]
Since $G$ is periodic, it holds that $2^{2}\rho_{n-2} \in \dZ$ by Lemma \ref{lem of rho}. Thus, it necessarily holds that
\[  \left( \frac{1}{\deg{v_{1}}}+\frac{1}{\deg{w_{1}}}\right) \in \dZ. \]
This implies that $\deg{v_{1}}=\deg{w_{1}}=1$ or $2$.
\end{proof}

If $\deg{v_{1}}=\deg{w_{1}}=1$, then $G$ is a member of the graphs seen in Figure \ref{even graph}, which are not odd-periodic by Theorem \ref{even period}. Thus, we suppose $\deg{v_{1}}=\deg{w_{1}}=2$. Here, we recursively define the vertices $v_{t} \in V(S_{1})$ and $w_{t} \in V(S_{2})$ for $t \in \dN$ as follows: If $\deg{v_{1}}=\dots=\deg{w_{t-1}}=2$, and $\deg{w_{1}}=\dots =\deg{w_{t-1}}=2$, then we define
\begin{align}
v_{t}&:=\{v \in V(S_{1}) \mid d(u,v)=t \} \label{def of vt}\\
w_{t}&:=\{w \in V(S_{2}) \mid d(u,w)=t\}. \label{def of wt}
\end{align}
\begin{figure}[h]
\begin{center}
\begin{tikzpicture}
[scale = 1.0,
fat/.style={circle,fill=black, inner sep = 2mm},
slim/.style={circle,fill=black, inner sep = 0.8mm},
]
\foreach \angle /\label in
{0,1,..., 8}
{
\node[slim] (v\angle) at (90+40*\angle:2) {};
\draw[line width=1pt] (90+40*\angle:2) -- (90+40*\angle+40:2);
} 
\node at (0,0) {$C_{k}$};
\node[slim] (x1) at (1,3) {};
\node[slim] (x2) at (2,4) {};
\node[slim] (x3) at (3,5) {};
\node[slim] (x4) at (3.75,5.5) {};
\node[slim] (x5) at (4,5) {};
\node[slim] (x6) at (3.5,6) {};

\node[slim] (y1) at (-1,3) {};
\node[slim] (y2) at (-2,4) {};
\node[slim] (y3) at (-3,5) {};
\node[slim] (y4) at (-3.75,5.5) {};
\node[slim] (y5) at (-4,5) {};
\node[slim] (y6) at (-3.5,6) {};
\draw[line width=1pt] (v0)--(x1);
\draw[line width=1pt, dashed] (1.2,3.2)--(1.8,3.8);
\draw[line width=1pt, dashed] (3.97,5.6)--(4.97,6);
\draw[line width=1pt, dashed] (4.3,5)--(5.3,5);
\draw[line width=1pt, dashed] (3.7,6.2)--(4.5,7.3);
\draw[line width=1pt] (x2)--(x3);
\draw[line width=1pt] (x3)--(x4);
\draw[line width=1pt] (x3)--(x5);
\draw[line width=1pt] (x3)--(x6);

\draw[line width=1pt] (v0)--(y1);
\draw[line width=1pt, dashed] (-1.2,3.2)--(-1.8,3.8);
\draw[line width=1pt, dashed] (-3.97,5.6)--(-4.97,6);
\draw[line width=1pt, dashed] (-4.3,5)--(-5.3,5);
\draw[line width=1pt, dashed] (-3.7,6.2)--(-4.5,7.3);
\draw[line width=1pt] (y2)--(y3);
\draw[line width=1pt] (y3)--(y4);
\draw[line width=1pt] (y3)--(y5);
\draw[line width=1pt] (y3)--(y6);
\node at (1.3,2.7) {$w_{1}$};
\node at (2.3,3.7) {$w_{t-1}$};
\node at (3.3,4.7) {$w_{t}$};

\node at (-1.3,2.7) {$v_{1}$};
\node at (-2.3,3.7) {$v_{t-1}$};
\node at (-3.3,4.7) {$v_{t}$};
\node at (0,2.3) {$u$};

\end{tikzpicture}
\caption{The setting of $v_{t}$ and $w_{t}$.}
\label{odd setting}
\end{center}
\end{figure}
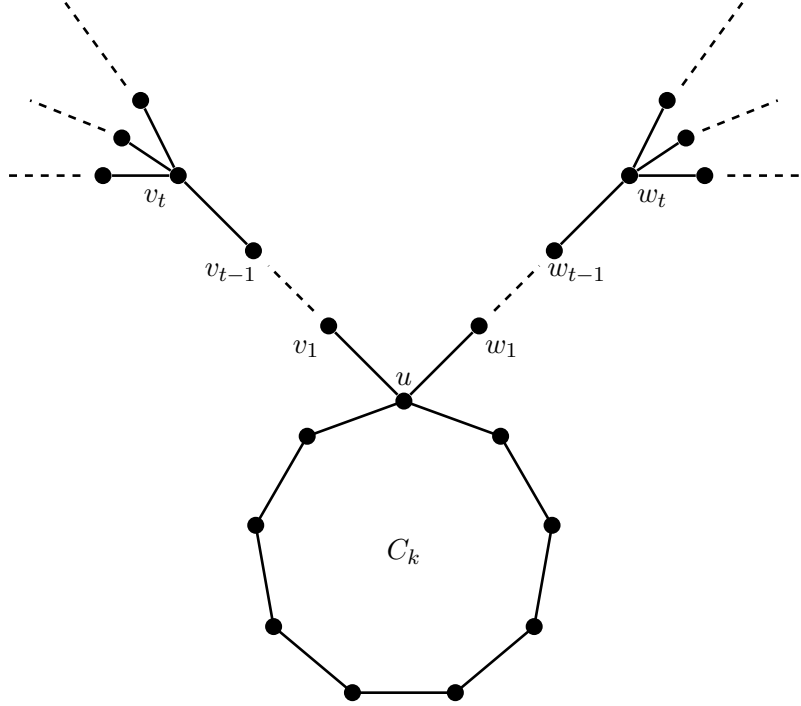
In this situation, the vertices are uniquely determined (See Figure \ref{odd setting}).  For $1 \le i \le t$, we set edges $g_{i}$ and $h_{i}$ using $g_{i}=v_{i-1}v_{i}$ and $h_{i}=w_{i-1}w_{i}$, respectively. We define 
\begin{align*}
K^{g}_{2i}(2, \dots, r)&:=\sum_{e_{1}, \dots e_{i} \in E(S)\setminus\{g_{2}, \dots, g_{r}\}}\prod^{i}_{l=1}M(e_{l}),\\
K^{h}_{2i}(2, \dots, r)&:=\sum_{e_{1}, \dots e_{i} \in E(S)\setminus\{h_{2}, \dots, h_{r}\}}\prod^{i}_{l=1}M(e_{l}),
\end{align*}
and
\[ L_{2i}(2, \dots, r):=K^{g}_{2i}(2, \dots, r)+K^{h}_{2i}(2, \dots, r) \]
for $1 \le r \le t-1$. It is convenient to set $K^{g}_{0}(2, \dots, r)=K^{h}_{0}(2, \dots, r)=1$ and $L_{0}(2, \dots, r)=2$. We next introduce a condition that the above values satisfy. 
\begin{lem}
Suppose $\deg{v_{1}}=\deg{v_{2}}=\dots=\deg{v_{t-1}}=2$ and $\deg{w_{1}}=\deg{w_{2}}=\dots=\deg{w_{t-1}}=2$. Let $K_{2i}$ and $L_{2i}(2, \dots r)$ be defined as above. For $1 \le i \le t$ and $2 \le r  \le t-1$,  it holds that
\[ L_{2i}(2, \dots, r)=2K_{2i}-\sum^{r}_{j=2}\left( \frac{1}{4}L_{2(i-1)}(2, \dots, j+1)\right). \]
\label{K and L}
\end{lem}
\begin{proof}
Recall that $K^{s}_{2i}(2, \dots, r)=\sum_{e_{1}, \dots, e_{i} \in E(S)\backslash\{s_{2}, \dots, s_{r}\}}\prod^{i}_{l=1}M(e_{l})$ for $s \in \{g, h\}$. Define $\sum^{j}_{e_{1}, \dots, e_{i}}$ as the summation over the combinations of edges in $E(S) \backslash \{s_{2}, \dots, s_{j+1}\}$, which form an $i$-matching containing $s_{j}$. Then it holds that
\begin{align*}
K^{s}_{2i}(2, \dots r)&=\sum_{e_{1}, \dots, e_{i} \in E(S)\backslash\{s_{2} \dots s_{r}\}}\prod^{i}_{l=1}M(e_{l})\\
&=\sum_{e_{1}, \dots, e_{i} \in E(S) }\prod^{i}_{l=1}M(e_{l})-\sum^{r}_{j=2}{\sum}^{j}_{e_{1}, \dots, e_{i}}\prod^{i}_{l=1}M(e_{l})\\
&=K_{2i}-\sum^{r}_{j=2}{\sum}^{j}_{e_{1}, \dots, e_{i}}\prod^{i}_{l=1}M(e_{l}).
\end{align*}
Here, we have
\begin{align*}
{\sum}^{j}_{e_{1}, \dots e_{i}}\prod^{i}_{l=1}M(e_{l})&=M(s_{j})\sum_{e_{1}, \dots, e_{i-1}\in E(S)\backslash\{s_{2}, \dots, s_{j+1} \}}\prod^{i-1}_{l=1}M(e_{l})\\
&=\frac{1}{4}K^{s}_{2(i-1)}(2, \dots, j+1).
\end{align*}
Therefore, it holds that
\[ K^{s}_{2i}(2, \dots, r)=K_{2i}-\sum^{r}_{j=2}\left( \frac{1}{4}K^{s}_{2(i-1)}(2, \dots, j+1)\right)\]
and 
\begin{align*}
L_{2i}(2, \dots, r)&=K^{g}_{2i}(2, \dots, r)+K^{h}_{2i}(2, \dots, r)\\
&=2K_{2i}-\sum^{r}_{j=2}\left( \frac{1}{4}L_{2(i-1)}(2, \dots, j+1)\right).
\end{align*}

\end{proof}
\begin{lem}
Let $K_{2t}$ be defined as above. It holds that
\[ K_{2t}=\rho_{n-2t}-\sum^{t-1}_{j=1}(A^{(t)}_{j}+B^{(t)}_{j}+C^{(t)}_{j}), \]
where 
\begin{align*}
A^{(t)}_{j}&=\left(\frac{1}{2}\right)^{2(t-j)}K_{2j}a^{(t)}_{j},\\
B^{(t)}_{j}&=\left(\frac{1}{2}\right)^{2(t-j)}K_{2j}b^{(t)}_{j},\\
C^{(t)}_{j}&=\left(\frac{1}{2}\right)^{2(t-j)+1}L_{2j}(2)c^{(t)}_{j},\\
\end{align*}
for some constants $a^{(t)}_{j}, b^{(t)}_{j}, c^{(t)}_{j} \in \dZ$. 
\label{lem of Kt}
\end{lem}
\begin{proof}
Recall that 
\[ K_{2t}=\sum_{e_{1}, \dots, e_{t} \in E(S)}\prod^{t}_{l=1}M(e_{l}).\]
Define ${\sum}'_{e_{1}, \dots, e_{i}}$ to be the summation over the combinations of edges in $E(G)$ forming an $i$-matching containing at least one edge in $E(G')$. Then we have
\begin{align*}
K_{2t}&=\sum_{e_{1}, \dots, e_{t} \in E(S)}\prod^{t}_{l=1}M(e_{l})\\
&=\sum_{e_{1}, \dots, e_{t} \in E(G)}\prod^{t}_{l=1}M(e_{l})-{\sum}'_{e_{1}, \dots, e_{t}}\prod^{t}_{l=1}M(e_{l})\\
&=\rho_{n-2t}-{\sum}'_{e_{1}, \dots, e_{t}}\prod^{t}_{l=1}M(e_{l}).
\end{align*}
Let $X_{t}={\sum}'_{e_{1}, \dots, e_{t}}\prod^{t}_{l=1}M(e_{l})$. $X_{t}$ is constructed by $t$-matchings containing at least one edge in $E(G')$. Suppose that there is a $t$-matching with $j$ edges in $E(S)$ and $(t-j)$ edges in $E(G')$ for $1 \le j \le t-1$. The candidates for such a $(t-j)$ matching given by the $(t-j)$ edges in $E(G')$ are separated into three cases:
\begin{itemize}
\item[(A-$j$)]: It has no edges from $E_{u}$,
\item[(B-$j$)]: It has $f_{1}$ or $f_{2}$,
\item[(C-$j$)]: It has $g_{1}$ or $h_{1}$. 
\end{itemize}  
In the case of (A-$j$), the $t$-matching is constructed using $(t-j)$ edges in $E(G')\backslash E_{u}$ and $j$ edges in $E(S)$, which yield 
\[  \sum_{e_{1}, \dots, e_{t-j} \in E(G')\backslash E_{u}}\prod^{t-j}_{i=1}M(e_{i}), \]
and
\[ \sum_{e_{1}, \dots, e_{j} \in E(S) }\prod^{j}_{l=1}M(e_{l})=K_{2j},\]
respectively. Denote the product of them as $A^{(t)}_{j}$. $A^{(t)}_{j}$ is a term contributing to $X_{t}$, which is constructed from $t$-matchings in $E(S)$ whose $(t-j)$ edges satisfy (A-$j$). 
For $e \in E(G')\backslash E_{u}$, it holds that $M(e)=\frac{1}{4}$. For $e_{1}, \dots, e_{t-j} \in E(G')\backslash E_{u}$, we thus have
\[ \prod^{t-j}_{i=1}M(e_{i})=\left(\frac{1}{2}\right)^{2(t-j)}.  \]
Note that this value does not depend on the choice of $(t-j)$-matchings. Denote the number of these $(t-j)$-matchings as $a^{(t)}_{j}$. If there is no such matching in $E(G')\backslash E_{u}$, then we set $a^{(t)}_{j}=0$. Thus, we have
\[ A^{(t)}_{j}=\left(\frac{1}{2}\right)^{2(t-j)}K_{2j}a^{(t)}_{j}. \]

We next consider the case of (B-$j$). Here, the value
\begin{align*}
M(f_{1}) &  \sum_{e_{1}, \dots, e_{t-j-1} \in E(G')\backslash E_{u}}\prod^{t-j-1}_{i=1}M(e_{i}) \sum_{e_{1}, \dots, e_{j} \in E(S) }\prod^{j}_{l=1}M(e_{l})\\
&=\frac{1}{8}\sum_{e_{1}, \dots, e_{t-j-1} \in E(G')\backslash E_{u}}\prod^{t-j-1}_{i=1}M(e_{i})K_{2j} 
\end{align*}
is constructed from $t$-matchings with $f_{1}$ and $(t-j-1)$ edges in $E(G')\backslash E_{u}$ and $j$ edges in $E(S)$. 
Similarly, $t$-matchings for (B-$j$) with $f_{2}$ yield the same value. 
Let
\[ 
B^{(t)}_{j}=\frac{1}{4}\sum_{e_{1}, \dots, e_{t-j-1} \in E(G')\backslash E_{u}}\prod^{t-j-1}_{i=1}M(e_{i})K_{2j}
\]
for $1 \le j \le t-2$ and $B^{(t)}_{t-1}=\frac{1}{4}K_{2(t-1)}$. This is a term contributing to $X_{t}$, which is constructed by $t$-matchings whose $(t-j)$ edges satisfy (B-$j$). 
It similarly holds that $M(e)=\frac{1}{4}$ for $e \in E(G') \backslash E_{u}$, and we have
\[ \prod^{t-j-1}_{i=1}M(e_{i})=\left(\frac{1}{2} \right)^{2(t-j-1)} \]
for $e_{1}, \dots, e_{t-j-1} \in E(G')\backslash E_{u}$. This does not depend on the choice of $(t-j-1)$-matchings. Denoting the number of these $(t-j-1)$-matchings as $b^{(t)}_{j}$, we have
\[ B^{(t)}_{j}=\frac{1}{4}\left(\frac{1}{2} \right)^{2(t-j-1)} K_{2j}b^{(t)}_{j}=\left(\frac{1}{2} \right)^{2(t-j)}K_{2j}b^{(t)}_{j}. \]

Finally, we consider the case of (C-$j$). A $t$-matching for this case is constructed using $g_{1}$, $(t-j-1)$ edges in $E(G') \backslash E_{u}$, and $j$ edges in $E(S) \backslash\{g_{2}\}$. Such a matching yields
\begin{align*}
M(g_{1})& \sum_{e_{1}, \dots, e_{t-j-1} \in E(G')\backslash E_{u}}\prod^{t-j-1}_{i=1}M(e_{i}) \sum_{e_{1}, \dots, e_{j} \in E(S)\backslash\{ g_{2} \} }\prod^{j}_{l=1}M(e_{l})\\
&=\frac{1}{8}\sum_{e_{1}, \dots, e_{t-j-1} \in E(G')\backslash E_{u}}\left( \frac{1}{4}\right)^{t-j-1}K^{g}_{2j}(2)\\
&=\sum_{e_{1}, \dots, e_{t-j-1} \in E(G')\backslash E_{u}}\left( \frac{1}{2}\right)^{2(t-j)+1}K^{g}_{2j}(2).
\end{align*}
A $t$-matching in this case with $h_{1}$ similarly yields 
\[
\sum_{e_{1}, \dots, e_{t-j-1} \in E(G')\backslash E_{u}}\left( \frac{1}{2}\right)^{2(t-j)+1}K^{h}_{2j}(2).
\]
Let
\begin{align*}
C^{(t)}_{j}&=\sum_{e_{1}, \dots, e_{t-j-1} \in E(G')\backslash E_{u}}\left( \frac{1}{2}\right)^{2(t-j)+1}K^{g}_{2j}(2)+\sum_{e_{1}, \dots, e_{t-j-1} \in E(G')\backslash E_{u}}\left( \frac{1}{2}\right)^{2(t-j)+1}K^{h}_{2j}(2)\\
&=\sum_{e_{1}, \dots, e_{t-j-1} \in E(G')\backslash E_{u}}\left( \frac{1}{2}\right)^{2(t-j)+1}L_{2j}(2)
\end{align*}
for $1 \le j \le t-2$ and $C^{(t)}_{t-1}=\frac{1}{8}L_{2j}(2)$. The value in this summation does not depend on the choice of $(t-j-1)$-matching as was the case above. Denoting the number of these $(t-j-1)$-matchings as $c^{(t)}_{j}$, we have
\[ C^{(t)}_{j}=\left( \frac{1}{2}\right)^{2(t-j)+1}L_{2j}(2)c^{(t)}_{j}, \]
which is the final term contributing to $X_{t}$. 

Therefore, we have
\[ X_{t}=\sum^{t-1}_{j=1}(A^{(t)}_{j}+B^{(t)}_{j}+C^{(t)}_{j}) \]
and 
\begin{align*}
K_{2t}&=\rho_{n-2t}-X_{t}\\
&=\rho_{n-2t}-\sum^{t-1}_{j}(A^{(t)}_{j}+B^{(t)}_{j}+C^{(t)}_{j}).
\end{align*}


\end{proof}

Now, we give key statements needed to prove our main result. 
\begin{thm}
Suppose that $G$ is periodic, $\deg{v_{1}}=\dots=\deg{v_{t}}=2$, and $\deg{w_{1}}=\dots=\deg{w_{t}}=2$. For $1 \le i \le t$, we have 
\begin{eqnarray}
2^{2i}K_{2i} \in \dZ, & \label{eq1}\\
2^{2(i-1)-1}L_{2(i-1)}(2) \in \dZ, & \label{eq2}\\
2^{2(i-j)-1}L_{2(i-j)}(2, \dots, j+1) \in \dZ, & 2 \le j \le i-1.\label{eq4}
\end{eqnarray}
\label{main thm1}
\end{thm}
\begin{proof}
We demonstrate this using induction on $i$. The case of $t \le 2$ is clear, so we turn our attention to $t>2$. Suppose that (\ref{eq1}), (\ref{eq2}), and (\ref{eq4}) hold for $1 \le i \le t-1$. We first prove that (\ref{eq4}) holds for $i=t$ and every $j$ with $2 \le j \le t-1$. We begin with the case of $i=t$ and $j=t-1$. Here, we have
\begin{align*}
K^{g}_{2}(2, \dots, t)&=\sum_{e \in E(S) \backslash\{ g_{2}, \dots, g_{t}\}}M(e)\\
&=K_{2}-\sum^{t}_{j=2}M(g_{j})\\
&=K_{2}-\frac{1}{4}(t-1).
\end{align*}
Similarly, we have $K^{h}_{2}(2, \dots, t)=K_{2}-\frac{1}{4}(t-1)$. Hence, it holds that
\[ L_{2}(2, \dots, t)=K^{g}_{2}(2, \dots, t)+K^{h}_{2}(2, \dots, t)=2K_{2}-\frac{1}{2}(t-1), \]
and this gives rise to 
\[ 2L_{2}(2, \dots, t)=2^{2}K_{2}-(t-1) \in \dZ \]
by the assumption in (\ref{eq1}). Then (\ref{eq4}) holds for $i=t$ and $j=t-1$. Using $i=t$ and $j=t-2$ in (\ref{eq4}), we have 
\[ L_{4}(2, \dots, t-1)=2K_{4}-\frac{1}{4}\sum^{t-1}_{j=2}L_{2}(2, \dots, j+1)\]
by Lemma \ref{K and L}. Then
\[ 2^{3}L_{4}(2, \dots, t-1)=2^{4}K_{4}-2\sum^{t-1}_{j=2}L_{2}(2, \dots, j+1). \]
Now, it holds that $2^{4}K_{4} \in \dZ$ and $2L_{2}(2, \dots, j+1) \in \dZ$ for $2 \le j \le t-2$ by the assumption. Additionally, it holds that $2L_{2}(2, \dots, t) \in \dZ$ by the previous argument, and this gives rise to $2^{3}L_{4}(2, \dots, t-1) \in \dZ$. Taking $i=t$ and $j=t-k$ for $3 \le k \le t-2$, we can show inductively that 
\[ 2^{2k-1}L_{2k}(2, \dots, t-k+1) \in \dZ. \]
Hence, we have
\[ 2^{2(t-j)-1}L_{2(t-j)}(2, \dots, j+1) \in \dZ \]
for $2 \le j \le t-1$. Thus, (\ref{eq4}) holds for $i=t$ and $2 \le j \le t-1$.  

We next show that (\ref{eq2}) holds for $i=t$. By Lemma \ref{K and L}, we have
\[ L_{2(t-1)}(2)=2K_{2i}-\frac{1}{4}L_{2(t-2)}(2,3). \]
Therefore, it holds that
\[ 2^{2(t-1)-1}L_{2(t-1)}(2)=2^{2(t-1)}K_{2(t-1)}-2^{2(t-2)-1}L_{2(t-2)}(2,3). \]
It follows from the assumptions for the case of $i=t-1$ in (\ref{eq1}) and (\ref{eq2}) that $2^{2(t-1)}K_{2(t-1)} \in \dZ$ and $2^{2(t-2)-1}L_{2(t-2)}(2, 3) \in \dZ$. Thus, (\ref{eq2}) holds for $i=t$. 

Finally, we show that (\ref{eq1}) holds for $i=t$. By Lemma \ref{lem of Kt}, it holds that
\[ K_{2t}=\rho_{n-2t}-\sum^{t-1}_{j=1}(A^{(t)}_{j}+B^{(t)}_{j}+C^{(t)}_{j}), \]
where $A^{(t)}_{j}, B^{(t)}_{j}$, and $C^{(t)}_{j}$ are defined as in the statement in Lemma \ref{lem of Kt}. Now, it holds that $2^{2t}\rho_{n-2t} \in \dZ$ by Lemma \ref{lem of rho}. Additionally, we have
\begin{align*}
2^{2t}A^{(t)}_{j}&=2^{2j}K_{2j}a^{(t)}_{j}\\
2^{2t}B^{(t)}_{j}&=2^{2j}K_{2j}b^{(t)}_{j}\\
2^{2t}C^{(t)}_{j}&=2^{2j-1}L_{2j}(2)c^{(t)}_{j}
\end{align*}
for $1 \le j \le t-1$. By the assumption, we have $2^{2j}K_{2j} \in \dZ$ for $1 \le j \le t-1$. Moreover, it follows from the assumption and the above argument that $2^{2j-1}L_{2j}(2) \in \dZ$ for $1 \le j \le t-1$. Hence, we have
\[ 2^{2t}\sum^{t-1}_{j=1}(A^{(t)}_{j}+B^{(t)}_{j}+C^{(t)}_{j}) \in \dZ \],
and this gives rise to
\[ 2^{2t}K_{2t} \in \dZ. \]
Therefore, we have completed the proof. 
\end{proof}

\begin{thm}
Suppose that $G$ is periodic, $\deg{v_{1}}=\dots=\deg{v_{t}}=2$, and $\deg{w_{1}}=\dots=\deg{w_{t}}=2$. We have
\[ \deg{v_{t+1}}=\deg{w_{t+1}}=1\quad \text{or}\quad 2. \]
\label{main thm2}
\end{thm}
\begin{proof}
Recall that the numbers $n$ and $k$ are the number of vertices of $G$ and the length of the unique cycle in $G$, respectively. Since $G$ is periodic, it holds that $2^{k+2(t+1)}\rho_{n-k-2(t+1)} \in \dZ$, and it follows from Lemma \ref{lem of K} that $2^{2(t+1)}K_{2(t+1)} \in \dZ$. By Lemma \ref{lem of Kt}, we have
\[ K_{2(t+1)}=\rho_{n-2(t+1)}-\sum^{t}_{j=1}(A^{(t+1)}_{j}+B^{(t+1)}_{j}+C^{(t+1)}_{j}). \]
Recall that $2^{2(t+1)}\rho_{n-2(t+1)} \in \dZ$. As is seen in the proof of Theorem \ref{main thm1}, we have 
\[ 2^{2(t+1)}A^{(t+1)}_{j}, 2^{2(t+1)}B^{(t+1)}_{j} \in \dZ \]
for $1 \le j \le t$. Thus, the condition of $2^{2(t+1)}K_{2(t+1)} \in \dZ$ is reduced to
\begin{equation}
2^{2(t+1)}\sum^{t}_{j=1}C^{(t+1)}_{j} \in \dZ. 
\label{final0}
\end{equation}
Now, recall that
\[ C^{(t+1)}_{j}=\left( \frac{1}{2}\right)^{2(t+1-j)+1}L_{2j}(2)c^{(t+1)}_{j}, \]
where $c^{(t+1)}_{j}$ is the number of $(t+1-j)$-matchings in $E(G') \cup \{g_{1}, h_{1}\}$ with $g_{1}$, which coincides with those with $h_{1}$. Thus, we have
\[ 2^{2(t+1)}C^{(t+1)}_{j}=2^{2j-1}L_{2j}(2)c^{(t+1)}_{j}.  \]
By Theorem \ref{main thm1} (\ref{eq2}), it holds that  
\[ 2^{2j-1}L_{2j}(2) \in \dZ \]
for $1 \le j \le t-1$. The condition in (\ref{final0}) then implies 
\[ 2^{2t-1}L_{2t}(2)c^{(t+1)}_{t} \in \dZ.  \]
Note that $c^{(t+1)}_{t}=1$ since the number of $1$-matchings in $E(G') \cup \{ g_{1}, h_{1}\}$ with $g_{1}$ is one. It suffices to consider
\begin{equation}
2^{2t-1}L_{2t}(2) \in \dZ. 
\label{final1}
\end{equation}
By Lemma \ref{K and L}, we have
\[ L_{2t}(2)=2K_{2t}-\frac{1}{4}L_{2(t-1)}(2,3) \]
and
\[ 2^{2t-1}L_{2t}(2)=2^{2t}K_{2t}-2^{2(t-1)-1}L_{2(t-1)}(2,3). \]
By Theorem \ref{main thm1} (\ref{eq1}), we have 
\[ 2^{2t}K_{2t} \in \dZ \],
and the condition in (\ref{final1}) is reduced to
\begin{equation}
2^{2(t-1)-1}L_{2(t-1)}(2,3) \in \dZ. 
\label{final2}
\end{equation}
By Lemma \ref{K and L}, we have
\[ L_{2(t-1)}(2,3)=2K_{2(t-1)}-\frac{1}{4}L_{2(t-2)}(2,3)-\frac{1}{4}L_{2(t-2)}(2,3,4).  \]
Then
\[ 2^{2(t-1)-1}L_{2(t-1)}(2,3)=2^{2(t-1)}K_{2(t-1)}-2^{2(t-2)-1}L_{2(t-2)}(2,3)-2^{2(t-2)-1}L_{2(t-2)}(2,3,4). \]
Since $2^{2(t-1)}K_{2(t-1)} \in \dZ$ and $2^{2(t-2)-1}L_{2(t-2)}(2,3) \in \dZ$ by Theorem \ref{main thm1} (\ref{eq1}) and (\ref{eq4}), the condition in (\ref{final2}) is reduced to
\begin{equation} 
2^{2(t-2)-1}L_{2(t-2)}(2,3,4) \in \dZ. 
\label{final3}
\end{equation}
Similarly, the condition in (\ref{final3}) implies
\[ 2^{2(t-3)-1}L_{2(t-3)}(2, 3,4,5) \in \dZ. \]
Repeating this process, we 
obtain the condition
\[ 2^{2(t+1-j)-1}L_{2(t+1-j)}(2, \dots, j+1) \in \dZ \],
which reduces to
\begin{equation}
2^{2(t-j)-1}L_{2(t-j)}(2, \dots, j+2) \in \dZ
\label{final4}
\end{equation}
for $4 \le j \le t-1$. Setting $j=t-1$ in (\ref{final4}), we obtain
\begin{equation}
2L_{2}(2, \dots, t+1) \in \dZ. 
\label{final5}
\end{equation}
On the other hand, we have
\begin{align*}
L_{2}(2, \dots, t+1)&=K^{g}_{2}(2, \dots, t+1)+K^{h}_{2}(2, \dots, t+1)\\
&=\sum_{e \in E(S)\backslash\{ g_{2}, \dots, g_{t+1}\}}M(e)+\sum_{e \in E(S)\backslash\{ h_{2}, \dots, h_{t+1}\}}M(e).\\
\end{align*}
Now,
\begin{align*}
\sum_{e \in E(S)\backslash\{ s_{2}, \dots,  s_{t+1}\}}M(e)&=\sum_{e \in E(S)}M(e)-\sum^{t+1}_{j=2}M(s_{j})\\
&=K_{2}-\frac{1}{4}(t-1)-\frac{1}{2\deg{v_{t+1}}}
\end{align*}
for $s \in \{g, h\}$. So, we have
\[ L_{2}(2, \dots, t+1)=2K_{2}-\frac{1}{2}(t-1)-\frac{1}{2}\left( \frac{1}{\deg{v_{t+1}}}+\frac{1}{\deg{w_{t+1}}}\right).  \]
Hence, the condition in (\ref{final5}) gives
\[ \frac{1}{\deg{v_{t+1}}}+\frac{1}{\deg{w_{t+1}}} \in \dZ, \]
which implies that $\deg{v_{t+1}}=\deg{w_{t+1}}=1$ or $2$. Thus, the proof is completed.

\end{proof}

\subsection{Proof of Theorem \ref{odd period}}
We will now finally prove Theorem \ref{odd period}. Let $G$ be an odd-periodic graph. Then it follows from Theorem \ref{odd cycle} that $G$ is an odd unicycle graph, that is, $G$ is a member of $\mathscr{O}(k)$ for an odd $k \in \dN$. Denote the vertices of the unique cycle in $G$ as $\{u_{0}, u_{1}, \dots, u_{k-1} \}$. By Theorem \ref{thm girth}, one of the following holds: 
\begin{itemize}
\item[(i)] $\deg{u_{0}}=\deg{u_{1}}=\dots=\deg{u_{k-1}}=2$, 
\item[(ii)] there exits $0 \le i \le k-1$ such that $\deg{u_{i}}=4$ and $\deg{u_{l}}=2$ for $l \ne i$.  
\end{itemize}
If (i) holds, then $G$ is nothing but an odd cycle. Suppose that (ii) holds and let $\deg{u_{0}}=4$ and $\deg{u_{i}}=2$ for $1 \le i \le k-1$. Thus, $G$ is the kind of graph seen in Figure \ref{odd setting}. Let two vertices $v_{1}$ and $w_{1}$ be defined as in this figure. In this case, it holds that $\deg{v_{1}}=\deg{w_{1}}=2$ by Theorem \ref{deg2}. Then we define two vertices $v_{2}$ and $w_{2}$ as in (\ref{def of vt}) and (\ref{def of wt}), respectively. Combining Theorems \ref{main thm1} and \ref{main thm2}, we have $\deg{v_{2}}=\deg{w_{2}}=1$ or $2$. If $\deg{v_{2}}=\deg{w_{2}}=2$, then we define two vertices $v_{3}$ and $w_{3}$ as in (\ref{def of vt}) and (\ref{def of wt}), respectively. Repeating this process, we have that if $\deg{v_{1}}=\deg{v_{2}}=\dots=\deg{v_{t}}=\deg{w_{1}}=\deg{w_{2}}=\dots=\deg{w_{t}}=2$ for some $t \in \dN$, then $\deg{v_{t+1}}=\deg{w_{t+1}}=1$ or $2$. Since $G$ is a finite graph, there exists $t \in \dN$ such that $\deg{v_{1}}=\deg{v_{2}}=\dots=\deg{v_{t}}=\deg{w_{1}}=\deg{w_{2}}=\dots=\deg{w_{t}}=2$ and $\deg{v_{t+1}}=\deg{w_{t+1}}=1$. This graph is nothing but the one illustrated in Figure \ref{even graph}. However, it is not odd periodic due to Theorem \ref{even period}. It follows then that odd-periodic odd unicycle graphs do not satisfy condition (ii). Therefore, the odd-periodic graph is only an odd cycle, and the proof is complete. 

\section{Summary and discussion}
In this paper, we tried to extract the graphs on which the induced Grover walk is odd-periodic, the odd-periodic graphs. It was previously observed that the odd-periodic graphs are included in a family of odd unicycle graphs (see Theorem \ref{odd cycle}). However, there are conditions that periodic odd unicycle graphs should obey (see Theorems \ref{thm girth}, \ref{deg2}, \ref{main thm1}, and \ref{main thm2}). These conditions are derived from the characteristic polynomial of the transition matrix defined in Section 2.3. Combining them, we see that the periodic odd unicycle graph illustrated in Figure \ref{even graph} is not an odd cycle, but the graph is not odd periodic. Therefore, the periodic odd unicycle graph whose period is odd is an odd cycle. In other words, the odd periodic graph is only an odd cycle. 

This result strongly depends on the combinatorial method used to compute the coefficients of the characteristic polynomial. We are currently focusing on counting combinations of cycles and matchings in the underlying graph to obtain the coefficients. Our ultimate goal is to control graph structure by utilizing properties of quantum walks. We believe that the work detailed in this paper helps us to reach that goal. 

By applying the above procedure, we can perfectly characterize the odd-periodic graphs. How to characterize the even-periodic graphs is still an open question, and it is one of our next research directions. 

\section*{Acknowledgement}
We would like to express our gratitude to Nobuaki Obata, Etsuo Segawa, Yusuke Higuchi, and Osamu Ogurisu for their continued support and fruitful comments. This study was supported by JSPS KAKENHI Grant Number 22K13952.

\end{document}